\documentclass[%
 reprint,
superscriptaddress,
 amsmath,amssymb,
 aps,
]{revtex4-2}

\usepackage{amsmath,amssymb,amsfonts,graphicx,url}
\graphicspath{{./},{./figures/}}

\newcommand{\argmin}{\mathop{\operatorname{argmin}}}

\usepackage{xcolor}
\usepackage{graphicx}
\usepackage{epstopdf}
\usepackage{graphicx}
\usepackage{dcolumn}
\usepackage{bm}

\begin{document}

\preprint{APS/123-QED}

\title{Inertialess Gyrating Engines}
\author{Jordi Ventura Siches}
\affiliation{%
Department of Mechanical and Aerospace Engineering, University of California, Irvine, CA 92697, USA
}
\author{Olga Movilla Miangolarra}
\affiliation{%
Department of Mechanical and Aerospace Engineering, University of California, Irvine, CA 92697, USA
}
\author{Amirhossein Taghvaei}
\affiliation{%
Aeronautics and Astronautics  Department, University of Washington, Seattle, WA 98195, USA
}%
\author{Yongxin Chen}
\affiliation{%
School of Aerospace Engineering, Georgia Institute of Technology, Atlanta, GA 30332, USA
}%
\author{Tryphon T. Georgiou}
\affiliation{%
Department of Mechanical and Aerospace Engineering, University of California, Irvine, CA 92697, USA
}%

\date{\today}
\begin{abstract}
A typical model for a gyrating engine 
consists of an inertial wheel powered by an energy source that generates an angle-dependent torque. 
Examples of such engines include a pendulum with an externally applied torque, Stirling engines, and the Brownian gyrating engine. Variations in the torque are averaged out by the inertia of the system to produce limit cycle oscillations.
While torque generating mechanisms are also ubiquitous in the biological world, where they typically feed on chemical gradients, inertia is not a property that one naturally associates with such processes.
In the present work, seeking ways to dispense of the need for inertial effects, we study an  
inertia-less concept where the combined effect of coupled torque-producing components averages out variations in the ambient potential and helps overcome dissipative forces to allow sustained operation for vanishingly small inertia.
We exemplify this inertia-less concept through analysis of two of the  aforementioned engines, the Stirling engine and the Brownian gyrating engine. An analogous principle may be sought in biomolecular processes as well as in modern-day technological engines, where for the latter, the coupled torque-producing components reduce vibrations that stem from the variability of the generated torque.
\end{abstract}

\maketitle


\section{Introduction}
The paradigm studied herein, referred to as a gyrating engine, is a system with a rotational degree of freedom characterized by an angle $\theta$ and driven by an external torque $\mathcal{T}$ that depends on $\theta$, which, however, may not necessarily retain the same sign during a cycle. Specifically, the device obeys
\begin{align}\label{eq:main}\vspace{-3pt}
\dot\theta &=\omega \nonumber \\
\mathcal I\dot\omega&=\mathcal T(\theta) -  \Gamma \omega,
\end{align}
where $\mathcal{I}$ is the moment of inertia and $\Gamma$ is the friction coefficient. We will refer to the term $-\Gamma\omega$ as external dissipation, though it could just as well represent torque proportional to angular velocity $\omega$ exchanged with an external subsystem acting as a load.
This model captures the general principle behind a wide range of mechanisms that convert thermal/chemical energy to rotary motion, whether synthetic or natural, from steam-engines to biomolecular motors.

We focus on two different types of gyrating engines, a low-temperature-differential Stirling engine \cite{izumida} that draws power from a temperature differential and a Brownian gyrating engine powered by Nyquist-Johnson thermal noise of two resistors kept at different temperature \cite{brownian_olga}. 
The salient feature in embodiments of these devices is the inertia needed to average out fluctuations and ensure sustained operation.
Analogous biomolecular mechanisms, however, seem to dispense of such a need for inertial effects \cite{meister1987proton,noji1997direct,kay2007synthetic}. A cursory view of the workings of biomolecular engines reveals a many-fold symmetry of multiple torque-generating units at work.
 With this in mind, we study the coupling of multiple gyrating engines as a way to eliminate the need for inertia in sustained limit cycle oscillation.

The basic idea explored in this paper is based on the principle that a {\em phase difference between coupled gyrating engines can average out the applied torque}. Thereby, angular variations in torque and load can be matched via a suitable geometric arrangement. We present analysis that highlights similarities between the two paradigms, the Stirling and Brownian gyrating engines, as well as provides quantitative and qualitative features of such arrangements. Our interest is mainly in enabling sustained operation in the presence of sign-indefinite generated torque by individual engines, that is, in ensuring that the combined torque of multiple units retains its sign.

The same principle can be used to minimize the variance of the effective torque being applied. Indeed, the idea of coupling engines to reduce torque variations is not new. Multi-cylinder internal combustion engines reduce torsional vibrations \cite{torquecancellarionIEEE,torsionalvibration}. However, exploring this principle for inertia-less operation of gyrating engines is new and may help elucidate the functionality of certain  biomolecular gyrating engines. 

Specifically, there are three motor proteins that have been unambiguously identified as rotary engines, the $F0/F1$ ATP synthase and the bacterial flagellar motor \cite{oster2003rotary}; they are powered by chemical gradients with the flagellar and $F0$ motor tapping onto trans-membrance ion-motive force while the $F1$ motor relying on ATP hydrolysis. Yet,
in spite of great strides over the past forty years into the workings of these $50nm$-scale motors, much remains to be understood \cite{sowa2008bacterial}. In regard to the mechanics, their geometry, that engages several torque-generating subunits
\cite{reid2006maximum,jia2019reconstitution} (up to $11$ in flagellar motors, and often a three-fold symmetry in ATPases), leads inescapably to the conclusion that a principle such as the one studied herein must be at work.

The structure of the paper is as follows. As part of the introduction, in Sections \ref{sec:A1} and \ref{sec:A2}, we present dynamical models for the Stirling engine and the Brownian gyrating engine. In Section \ref{sec:mainresults} we explain how a suitable geometry of a multi-engine coupled system operates without the need for inertia, and highlight the role of phase difference in sustaining operation as well as in optimizing other performance metrics. In Section \ref{sec:discussion} we summarize the gained insights. Finally, in the Appendixes, we provide proofs and expand on
technical statements given in the body of the paper.

\section{Examples of gyrating engines} \label{Examples}
We describe the two main paradigms of gyrating engines that are being considered along with their respective mathematical models.

\subsection{Stirling engine}\label{sec:A1}
\begin{figure}[h]
    \centering
    \includegraphics[width=0.325 \textwidth]{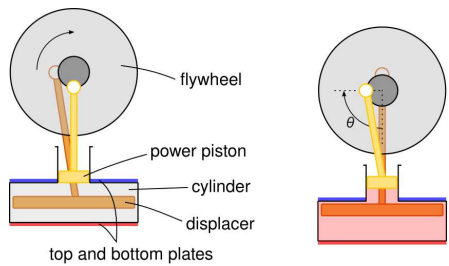}
    \caption{Parts of the Stirling engine and definition of angle $\theta$.}
    \label{fig:stirling_parts}
\end{figure}

The first gyrating engine that we consider is the so-called Stirling engine, invented by Robert Stirling in 1816, that generates mechanical work from a temperature differential. It consists of a cylinder filled with gas whose volume is adjusted by an oscillating piston -- the {\em power} piston -- connected to a flywheel with a slider-crank mechanism. Attached to this wheel and with a $\pi/2$ phase difference with respect to the power piston, there is another rod that is connected to a {\em displacer} piston, that forces the gas to switch sides and alternate contact with heat baths at the two sides, top and bottom plates, of the cylinder. Temperature fluctuations in the gas result in changes in the internal pressure, which drive the power piston accordingly (see Figure \ref{fig:stirling_parts}).
A detailed exposition along with simplified models for a typical Stirling engine have been presented recently in the timely work by Izumida and Toyabe \cite{Izumida_2018,izumida,izumida2020}.

In order for the engine to operate sustainably, the temperature difference must exceed a certain threshold, as noted in \cite{stirling_modeling}; we also refer to  \cite{stirling_molecular} for a detailed exposition of the coupling between the thermal gradient and the mechanics of the Stirling engine from a thermodynamic perspective.

Indeed, the underlying thermodynamics of the Stirling engine cycle have been thoroughly studied \cite{stirling_review,stirling_review2,stirling_review3,stirling_review4,organ}. However, models that include the gyrating dynamics of the engine are scarce. The simplified model that we adopt herein is based on the one developed in \cite{izumida} that has two degrees of freedom, the flywheel angle $\theta$ and its angular velocity $\omega=\dot{\theta}$. The equations of motion are those given in 
\eqref{eq:main} with the torque 
given by
\begin{equation}\label{eq:eom}
\mathcal T^{\rm S}(\theta)=sr(p(\theta)-p_0)\sin \theta,
\end{equation}
where $s$ is the section area of the power piston, $r$ is the crank radius, $p(\theta)$ is the pressure inside the cylinder and $p_0$ is the external atmospheric pressure. The pressure $p(\theta)$ is estimated using the ideal gas law scaled by a dimensionless parameter $\zeta$ that accounts for the nonuniformity of temperature and pressure in the cylinder, and it is
\begin{equation*}
    p(\theta)=\zeta\, \frac{nRT(\theta)}{V(\theta)},
\end{equation*}
where $n$ is the number of moles of gas in the cylinder and $R$ is the molar gas constant. The {\em effective} temperature $T(\theta)$ and the volume $V(\theta)$ of the gas in the cylinder can be expressed as follows,
\begin{equation*}
    \begin{split}
        T(\theta) &=T_{0}+\alpha  \frac{\Delta T}{2}\sin (\theta),\\
        V(\theta) &= V_0 + sr(1-\cos\theta),
    \end{split}
\end{equation*}
where $T_0 = (T_\text{top}+T_\text{btm})/2$ is the mean of the top and bottom temperatures, $\alpha$ is a dimensionless coefficient that models the heat transfer, 
$\Delta T = T_\text{btm} - T_\text{top}$ is the temperature difference and $V_0$ is the volume at $\theta=0$. 

We remark that in the model proposed by \cite{izumida}, the temperature is more generally expressed as a function of both $\theta$ and $\omega$. Specifically, the temperature's dependence on the angular position of the engine is delayed by a factor of $\tau \omega$, with $\sin(\theta-\omega\tau)$ replacing $\sin(\theta)$ in 
\eqref{eq:eom}. However, experimental evidence \cite{izumida} suggests that $\tau=15\times 10^{-3}$[sec]. Thus, in our analysis, we have adopted the simplifying assumption that $\omega\tau\simeq 0$; numerical simulations confirm that for our purposes, the effect of the small delay $\tau$ is indeed negligible.

\subsection{Brownian gyrating engine}\label{sec:A2}
The second example is that of a Brownian gyrator-based engine that was recently introduced in~\cite{brownian_olga}.
This consists of the coupling between an electrical system, known as the Brownian gyrator~\cite{brownian-filliger}, and a mechanical subsystem with an inertial wheel. Note that we distinguish between the Brownian gyrator and the Brownian gyrating engine, that consists of coupling the Brownian gyrator to the mechanical subsystem that mediates energy extraction.

\begin{figure}[t]
    \centering
    \includegraphics[width=0.35\textwidth]{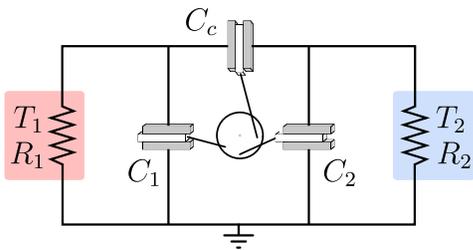}
    \caption{A physical embodiment of the Brownian gyrator consisting of an RC circuit.}
    \label{fig:BG}
\end{figure}

The electrical embodiment of the Brownian gyrator consists of three capacitors and two resistors (see Fig.~\ref{fig:BG}), which are in contact with two heat baths at different temperatures giving rise to Johnson-Nyquist fluctuating currents at the two resistors. This particular embodiment was introduced in \cite{brownian_circuit_PR}; equivalent realizations have been extensively studied, both theoretically \cite{brownian-filliger,Bgyrator2013ciliberto,BGyrator2013dotsenko,EnergyHarvesting2021} and experimentally~\cite{brownian_experimental,BGyrator2013CilibertoExperim,Bgyrator2013ciliberto}.

The mechanical subsystem includes dielectric padding in the three capacitors that can vary in its position through mechanical coupling to the rotating wheel as depicted in Fig.~\ref{fig:BG}. In this way, the angular position $\theta$ of the (inertial) wheel forces the dielectric material in and out of the respective capacitors. This mechanical coupling renders the capacitance-matrix a function of the dynamic variable $\theta$. In our analysis, the geometry of the linkages actuating the dielectric material has been chosen such that the capacitance matrix as a function of $\theta$ is of the form
\begin{equation*}
\begin{split}
 \small   C(\theta) &= \begin{bmatrix}
 C_1(\theta) + C_c(\theta) & -C_c(\theta)\\ - C_c(\theta) & C_2(\theta) + C_c(\theta)
 \end{bmatrix}\\
&= \hspace{-1pt}C_0\hspace{-1pt}\left[ \begin{array}{cc}
     \hspace{-4pt}2+\beta g_1(\theta) &  {- 1-\beta \cos(\theta)} \\
       {-1-\beta\cos(\theta)} & \hspace{-1pt}2+\beta g_2(\theta)
    \end{array}\hspace{-4pt}\right],
    \end{split}
\end{equation*}
where $C_1$, $C_2$ and $C_c$, depicted in Fig. \ref{fig:BG}, are expressed in terms of a nominal capacitance $C_0$, and the $\theta$-functions $g_1(\theta)=\cos(\theta+2\pi /3) +\cos(\theta)$ and $g_2(\theta)=\cos(\theta-2\pi /3) +\cos(\theta)$, with $0<\beta<1$.
The mechanical part can provide inertia as well as a resistive torque (modeled as $-\Gamma \frac{d\theta}{dt}$) that absorbs generated power.

As long as there is enough time-scale separation between the mechanical and the electrical subsystems, as shown in~\cite{brownian_olga}, the dynamics of the Brownian gyrating engine obey \eqref{eq:main} with 
\begin{equation*}
\begin{aligned}\mathcal T^{\rm B}(\theta)=-\frac{1}{2} \operatorname{Tr}\left[\partial_{\theta} C^{-1}(\theta) \Sigma(\theta)\right],
\end{aligned}
\end{equation*}
where $\text{Tr}[\cdot]$ denotes the trace operation, and $\Sigma (\theta)$ is the matrix covariance of the (Gaussian) state-vector $q_t = [q_1(t),q_2(t)]^\prime$  of charges at the two capacitors $C_1$ and $C_2$, respectively. By virtue of the time-scale separation, the matrix covariance satisfies the algebraic Lyapunov equation 
\begin{equation*}
    -R^{-1} C^{-1}(\theta) \Sigma(\theta)-\Sigma(\theta) C^{-1}(\theta) R^{-1}+R^{-1} D D^{\prime} R^{-1}=0,
\end{equation*}
with diffusion matrix $R = \text{diag}([R_1,R_2])$, $D = \text{diag}([\sqrt{2k_BR_1 T_1},\sqrt{2k_BR_2 T_2}])$, $k_B$ the Boltzmann constant and
$R_1$, $R_2$, $T_1$, $T_2$ as in Fig.~\ref{fig:BG}. The solution $\Sigma(\theta)$ of the above equation can be conveniently expressed compactly as a function of $C(\theta)$ as follows, 
\begin{equation*}
\Sigma(\theta)=\int_{0}^{\infty} e^{-R^{-1} C^{-1}(\theta) s} R^{-1} D D^{\prime} R^{-1} e^{-R^{-1} C^{-1}(\theta) s} d s.
\end{equation*}

\begin{figure}[b]
    \centering
    \includegraphics[width=0.45\textwidth]{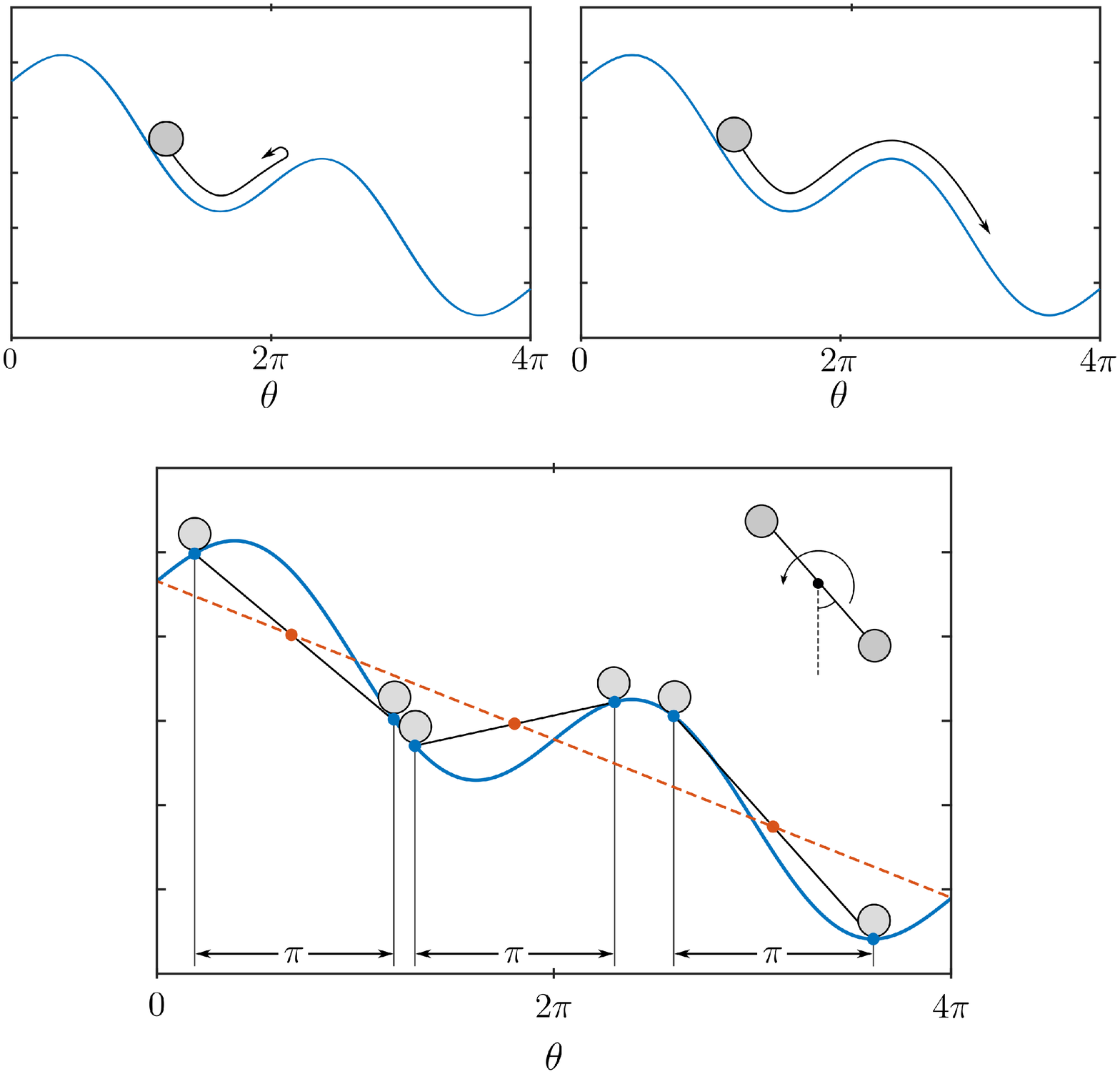}
    \caption{Potential for the damped pendulum with constant torque. Two cases are displayed. Top-left: inertial effects are not able to overcome the uphills generated by gravity and the only stable solution is the stationary one. Top-right: both inertial effects and constant torque (slope) are enough to sustain continuous motion and the pendulum reaches a stable periodic orbit. Bottom: the average of two potentials displaced by a $\pi$ phase difference is linear in $\theta$. The graphic representation provides insight into how two $\theta$-equispaced coupled pendula with a constant torque operate stably in a limit cycle: their combined effective potential is a sloped line (red dashed line in the figure). }
    \label{fig:pendulumpot}
\end{figure}
\subsection{Remark on the forced-pendulum abstraction}
It was noted in \cite{izumida,Izumida_2018,brownian_olga} that, in both examples, the resulting dynamical system's behaviour resembles that of the damped pendulum with constant torque \cite{damped_pendulum}, i.e., to a system that behaves according to \eqref{eq:main} with
\begin{equation*}
    \mathcal T^{\rm p}(\theta)=\gamma-\sin(\theta),
\end{equation*}
with $\gamma$ representing the constant torque being applied. It is insightful to consider the effective potential that drives the motion,
\begin{equation*}
    U(\theta)=- \int_0^\theta \mathcal T(\vartheta) d\vartheta.
\end{equation*}
This has the form of a tilted sinusoid.
A cartoon in two parts, corresponding to two different sets of parameters (of inertia and frictional forces), is displayed at the top row of Fig.~\ref{fig:pendulumpot}.
In this, the position of a ball rolling down the corrugated hill-side embodies the state of the pendulum; the drawing on the left
exemplifies insufficient-inertia/excessive-friction for a limit cycle to exist, while the one on the right exemplifies a continuous operation.
The second row of Fig.~\ref{fig:pendulumpot} depicts
the collaborative effect of two coupled engines. In the analogy of two coupled balls, the combined center of gravity is effectively lifted so as to facilitate sliding down the periodic potential which is tilted due to the applied torque.

The situation with the Stirling and Brownian gyrating engines is analogous. The coupling of a number of engines, with a suitable phase difference between one another, averages out the ``bumps'' in the ``corrugated'' potential and enables sustained operation for a vanishingly small applied torque.

\section{Results}
\label{sec:mainresults}
We begin by highlighting the effect of coupling several damped pendula with an applied constant torque and a certain phase difference. Specifically, for this case, we consider two pendula coupled with a phase difference of $\pi$ radians (see Fig.~\ref{fig:pendulumpot}, bottom). The effective torque on the combined system is
\begin{align*}
    \mathcal T^{\rm p}_2(\theta)&=\frac{1}{2}\left(\mathcal T^{\rm p}(\theta)+\mathcal T^{\rm p}(\theta+\pi)\right)\\&=\gamma-\frac{1}{2}\left(\sin(\theta)+\sin(\theta+\pi)\right)=\gamma,
\end{align*}
effectively canceling the undulations of the potential;
the $\frac12$ factor scales the power of the two engines so as that they can be compared to one engine. Thus, the effective torque remains constant, and thereby the overall
potential driving the system of two engines has a constant tilt with no undulations. The system requires neither any inertia nor a minimum amount of actuation to achieve sustained continuous rotation. The cartoon shown in Figure \ref{fig:pendulumpot} helps exemplify the effect.
The underlying principle is readily seen to rely on the cancellation of respective terms in a Fourier series expansion of the applied torque (see the Appendix \ref{bounds} for more details). Evidently, in more complicated examples, higher order harmonics are not immune and can likewise be eliminated or suppressed by coupling more engines as shown in the analysis that follows.

\begin{figure}[htb]
    \centering
    \includegraphics[width=0.475\textwidth]{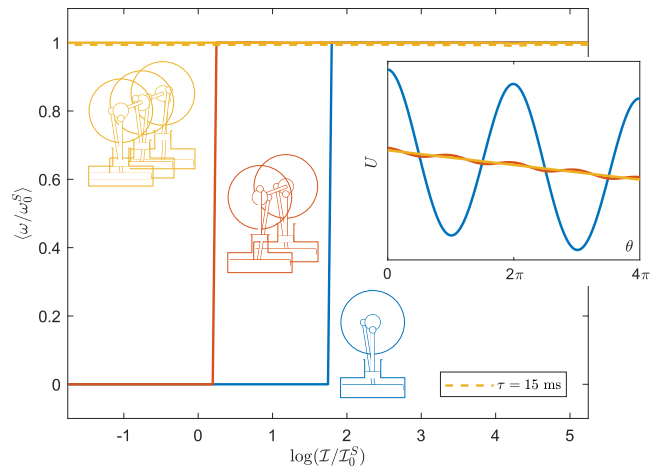}
    \caption{Left: normalized averaged steady state angular velocity $\langle\omega /\omega_0^S\rangle$ vs $\log(\mathcal{I}/\mathcal{I}_0^S)$ for one, two and three coupled engines, with $\Delta T = 10$K. Note that $\mathcal{I}$ is normalized by $\mathcal I_0^S=\Gamma/\omega_0^S$ and plotted in a logarithmic scale, where $\omega_0^S$ is  obtained from \eqref{eq:approximation_stirling}. Similarly, the angular velocity is also normalized by $\omega^S_0$. The case with $\tau=15$ ms is plotted in a dashed line and shows to what extent the assumption of the torque being $\omega$-independent holds. Right: effective potential along two cycles for one, two and three coupled engines.} 
    \label{fig:stirling_omegaIGamma123}
\end{figure}
\subsection{Inertialess Stirling engine}
We consider the equidistant (in the $\theta$ space) coupling of two and three Stirling engines, that generate combined torque
\begin{align*} \mathcal T^{\rm S}_2(\theta)&=\frac{1}{2}\left(\mathcal T^{\rm S}(\theta)+\mathcal T^{\rm S}(\theta+\pi)\right), \mbox{ and}\\
    \mathcal T^{\rm S}_3(\theta)&=\frac{1}{3}\left(\mathcal T^{\rm S}(\theta)+\mathcal T^{\rm S}(\theta+2\pi/3)+\mathcal T^{\rm S}(\theta+4\pi/3)\right),
\end{align*}
for the two- and three-engine configuration, respectively.
The resulting potential $U$ is shown in the insert in Fig.~\ref{fig:stirling_omegaIGamma123} over two periods. It changes from a periodic slopped shape (in the case of one engine), to practically a slopped straight line already for two coupled engines, and more so for three. The main plot in Fig.~\ref{fig:stirling_omegaIGamma123} shows the averaged steady state angular velocity as a function of inertia. It is seen that, for this set of parameters,
three engines dispense completely of the need for inertia, ensuring a limit cycle; with two engines the need for inertia is already minimal.

\begin{figure}[t]
    \centering \includegraphics[width=0.425\textwidth]{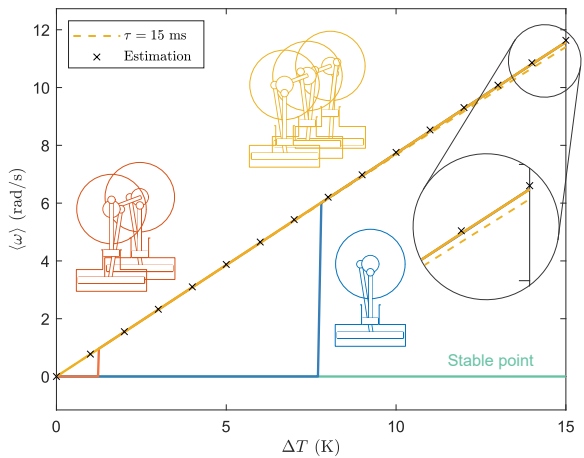}
    \caption{Averaged limit cycle angular velocity $\langle \omega \rangle$ as a function of the temperature difference $\Delta T$ for one, two and three coupled Stirling engines
    (solid lines). The yellow dashed line represents the case with $\tau=15$ms and three coupled engines, and numerically shows to what extent our assumption of the torque being $\omega$-independent is valid. This agreement is highlighted in the blow-up of the figure. An estimation of the average angular velocity in the limit cycle, based on \eqref{eq:approximation_stirling}, has been plotted in black crosses showing a good agreement with the numerical results.}
    \label{fig:stirling_DTomega_log2}
\end{figure}

Fig.~\ref{fig:stirling_DTomega_log2} illustrates how the averaged final angular velocity varies with the temperature difference $\Delta T$ that powers the engine(s). Evidently, the coupling of multiple Stirling engines reduces the threshold temperature difference needed for continuous operation. When three engines are coupled, the threshold is virtually eliminated, guaranteeing the existence of a limit cycle for vanishingly small $\Delta T$. In Appendix \ref{bound_stirling} we show that a sufficient condition for the torque $\mathcal T_3^S(\theta)$ to be always positive is
$$
\frac{\alpha \Delta T}{4T_0} \gg \frac{sr}{V_0}=:\epsilon,
$$
for a typical value $\epsilon \sim 10^{-3}$ in experimental settings of~\cite{izumida}.
The mean angular velocity of the wheel is, up to first order in $\epsilon$,
\begin{equation}\label{eq:approximation_stirling}
    \langle\omega\rangle\approx\frac{\alpha\Delta T \zeta n R}{4 \Gamma}\,\epsilon =:\omega_0^S,
\end{equation}
in complete agreement with the numerical results (see Appendix \ref{bounds} for the derivation). Also, note that the dependence of the angular velocity on the temperature difference is linear, confirming the hypothesis first introduced by Kolin~\cite{kolin} and experimentally supported by Toyabe and Izumida, and  Boutammachte and Norr \cite{izumida,boutammachte}. 

\subsection{Inertialess Brownian gyrating engine} 
\begin{figure}[t]
    \centering
    \includegraphics[width=0.4825\textwidth]{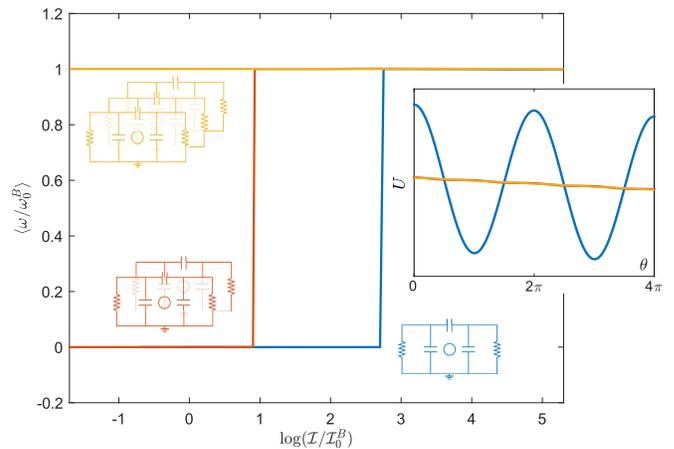}
    \caption{Left: normalized averaged final angular velocity $\langle \omega / \omega_0^B \rangle$ vs $\log{(\mathcal{I}/\mathcal{I}_0^B)}$, with $\Delta T = 10$K for one, two and three coupled Brownian gyrating engines, respectively. As before, $\mathcal{I}_0^B = \Gamma /\omega_0^B$ and $\omega_0^B$ is as defined in \eqref{eq:approximation_brownian}. Right: effective potential along two cycles.}
    \label{fig:brownian_omegaI_log2}
\end{figure}

We now consider the coupling of two and three Brownian gyrating engines with combined torque
\begin{align*} \mathcal T^{\rm B}_2(\theta)&=\frac{1}{2}\left(\mathcal T^{\rm B}(\theta)+\mathcal T^{\rm B}(\theta+\pi)\right), \ \mbox{and}\\
    \mathcal T^{\rm B}_3(\theta)&=\frac{1}{3}\left(\mathcal T^{\rm B}(\theta)+\mathcal T^{\rm S}(\theta+2\pi/3)+\mathcal T^{\rm B}(\theta+4\pi/3)\right),
\end{align*}
respectively. The resulting potential $U$ is drawn over two periods in the insert of Fig.~\ref{fig:brownian_omegaI_log2}. It displays the same qualitative behavior as that of the Stirling engine's potential. As we decrease the inertia we observe that the the limit cycle is similarly maintained in the case of three engines for vanishingly small inertia (see Fig.~\ref{fig:brownian_omegaI_log2}).

\begin{figure}[t]
    \centering
    \includegraphics[width=.45\textwidth]{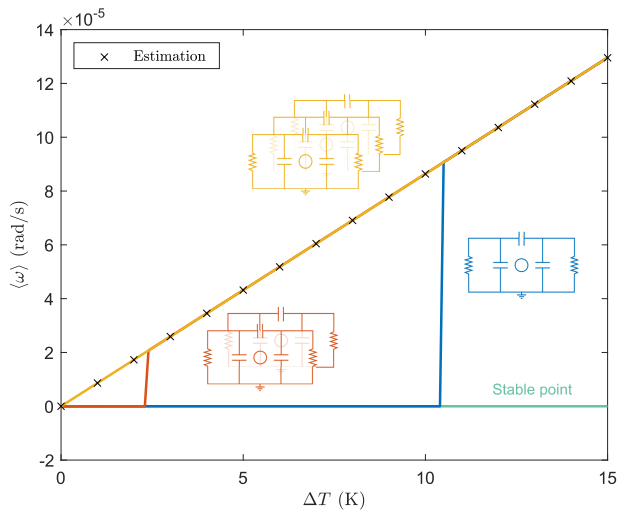}
    \caption{Average limit cycle angular velocity $\langle\omega\rangle$ vs. $\Delta T$ for one, two and three coupled Brownian gyrating engines. An estimation of the average angular velocity, from \eqref{eq:approximation_brownian}, has been plotted in black crosses, matching the numerical results.}
    \label{fig:brownian_DTomega_log2}
\end{figure}

Fig.~\ref{fig:brownian_DTomega_log2} displays the averaged angular velocity during operation as a linear function of the temperature difference  $\Delta T:=T_2-T_1$ that powers the gyrator, beyond a threshold that decreases with the number of coupled engines, as before. Similarly to the Stirling case, one can derive a sufficient condition for the existence of a limit cycle, namely,
$$
\frac{ \sqrt{3}}{64}\frac{\Delta T}{T_0}\gg \mathcal \beta
$$
where $T_0=(T_1+T_2)/2$. When this limit cycle is present, we can approximate the average angular velocity as
\begin{equation}\label{eq:approximation_brownian}
    \langle \omega \rangle  \approx \frac{\sqrt{3} k_B \Delta T}{64\Gamma}\, \beta^2 =: \omega_0^B,
\end{equation}
up to second order terms in $\beta$.

\subsection{Remarks on equalizing the torque}\label{sec:equalizing}
A main objective in coupling engines, in our exposition so far, has been the sustenance of inertialess operation. To this end we sought to cancel harmonics by coupling engines with equal phase difference from one another (equidistantly). However, this is by no means the only metric that one may adopt for quantifying performance. In particular, one may optimize the phase difference between engines as to maximize the minimal value of the torque along the cycle. Another possible metric for selecting phase differences is the variance of the torque, so as to limit vibrations. We highlight  this point by considering the special case of two engines, to be coupled accordingly.

We discuss the case where we seek to minimize the variance of the effective torque, in coupling two engines. That is, we seek
\begin{equation*}
  \theta_0^{\rm opt} =  \argmin_{\theta_0} \frac{1}{2\pi}\int_0^{2\pi} \left(\tfrac{\mathcal{T}(\theta) + \mathcal{T}(\theta + \theta_0)}{2}-\langle {\mathcal T}\rangle \right)^2d\theta,
\end{equation*} 
where $\langle{\mathcal T}\rangle$ is the mean value of the applied torque over a cycle and $\theta_0$ represents a phase difference between engines.

Clearly, $\theta_0=0$ maximizes the variance of the effective torque. As one may expect, $\theta_0=\pi$ represents another potential extremum. However, whether it corresponds to a minimum, a maximum or an inflection point depends on the specific shape of the torque-profile as a function of $\theta$. For instance, for a Stirling engine (keeping terms up to second order in $\epsilon=\tfrac{sr}{V_0}$), we obtain that, as long as $b_1^2<4a_2^2$, $\theta_0=\pi$ corresponds to a maximum while the minimum is achieved for
$$
\theta_0^{\rm opt}=\text{acos} \left(-\frac{b_1^2}{4a_2^2}\right),
$$
 where $b_1=\epsilon \, \left(1-\frac{p_0V_0}{\zeta n R T_0}\right)$ and $a_2=\epsilon \frac{\alpha \Delta T}{4 T_0}$ (see Appendix \ref{optimization}).
Otherwise, $\theta_0^{\rm opt}=\pi$, a value which was confirmed by our numerical experiments. Intuitively, $\pi$ is the optimal solution when the odd harmonics in $\mathcal T(\theta)$ dominate. The general case with a larger number of coupled engines can be worked out similarly.

\section{Conclusions} \label{sec:discussion}
The present paper details a proof-of-concept, that in gyrating engines, the need for inertia to ensure limit cycle oscillations can be dispensed of when a number of torque-generating sub-units are coupled with a suitable phase difference from one another.
When the effective torque produced by the combined contribution of subunits remains 
sign-definite over a cycle,
the system operates in a limit cycle making power available for external work. The underlying principle was demonstrated with two examples, a Stirling engine and a Brownian gyrating engine.

It is postulated that a similar principle is at work in biomolecular engines, albeit in a significantly more complicated guise, given the complexity of such engines.
Indeed, in \cite{mandadapu2015mechanics},
a model was presented and partially tested to explain specific physical
mechanisms for torque generation in bacterial flagellar motors (BFMs). In this, a number of torque generating units with a ``wide and gently slopping energy well'' contribute in ways that are reminiscent of the principle presented herein. Although the physics of torque generation remain poorly understood, it was proposed in \cite{mandadapu2015mechanics} that both electrostatic and steric forces are at work, with the latter generating a ``push''. The resulting torque profile may likely necessitate multiple units to smooth out higher harmonics that may thus be present. Understanding how ion-driven molecular machines work is of fundamental importance
in cellular biology, and thus the authors see likely that the principle discussed herein may help explain the workings of multiple torque-generating subunits
\cite{reid2006maximum,jia2019reconstitution} and, perhaps, even the necessity for a large number (up to $11$ in flagellar motors) of such units for the corresponding torque-generating potential.

\begin{acknowledgments}
OMM was supported by  ”la Caixa” Foundation (ID 100010434) with code LCF/BQ/AA20/11820047. The research was also supported in part by the NSF under grants 1807664, 1839441, 1901599, 1942523, and the AFOSR under FA9550-17-1-0435. JVS and OMM carried out the technical development of the work, OMM oversaw the completion, all authors, JVS, OMM, AT, YC and TTG contributed to the writing and development of ideas, TTG proposed the topic.
\end{acknowledgments}

\appendix
\label{sec:materials}

\section{Number of gyrating engines required to dispense of inertia}\label{bounds}
We consider gyrating engines obeying \eqref{eq:main}. Following two different approaches we show that provided the torque profile $\mathcal T(\theta)$  satisfies $
 |{\mathcal T}(\theta+\Delta)-{\mathcal T}(\theta)|<L|\Delta|$ for all $\theta,\Delta$ and with $L<\infty$ (i.e., it is Lipshitz) and provided the average torque over a cycle is not zero (without loss of generality, assumed positive), there is an integer $m$ so that $m$ equidistantly-coupled engines ensure a globally attractive limit cycle.
 In other words, we establish that under natural and mild conditions on the torque profile, a finite number of coupled Stirling or Brownian gyrating engines is always sufficient to maintain a stable limit cycle for \emph{any} set of parameters.

A sufficient condition for existence of a globally attractive limit cycle, regardless of $\mathcal{I}$, is that the torque $\mathcal{T}(\theta)$ is strictly positive. This is due to the Poincaré–Bendixson theorem,
which states that any trajectory in a bounded two-dimensional region of the phase plane, that contains no fixed points, must converge to a stable limit cycle. This applies to our case  as long as $\mathcal T(\theta)$ is never zero, precluding the existence of fixed points.

We now prove that a finite number of coupled engines ensures strictly positive torque for all values of the angular position $\theta$. Assuming that $\mathcal T(\theta)$ is Lipschitz and periodic, we consider the Fourier series expansion
\begin{equation}\label{eq:expansion}
    \mathcal{T}(\theta) = c_0 + \sum_{k=1}^\infty a_k\cos(k \theta) + b_k \sin(k\theta).
\end{equation}
 For $m$ equidistantly-coupled engines the effective torque is
 \footnotesize
\begin{align*}
    \mathcal{T}_m ( \theta ) &= \frac{1}{m} \sum_{\ell=0}^{m-1} \mathcal{T}\left(\theta -\tfrac{2\pi }{m} \, \ell\right) \\
    &  = c_0 + \frac{1}{m}\sum_{k=1}^\infty\sum_{\ell=0}^{m-1} a_k\cos(k(\theta-\tfrac{2\pi }{m} \, \ell)) + b_k \sin(k(\theta-\tfrac{2\pi }{m} \, \ell)) \\
    & = c_0 + \sum_{k=1}^\infty a_{km}\cos(km\theta) + b_{km} \sin(km\theta)\\
    &= \sum_{k=-\infty}^\infty c_{km} e^{ikm\theta},
\end{align*}
\normalsize where 
$c_k=\tfrac{1}{2}(a_k + i b_k)$ for $k>0$ and $c_k = \overline{c_{-k}}$ for $k<0$.
The third equality follows from cancellation, due to phase difference, of all terms with indices that are not multiples of $m$.
Since $\mathcal T(\theta)$ is Lipschitz, the amplitude of the harmonics decays faster than $k^{-1}$ and the series $\{|c_k| \mid k>0\}$ is summable, see e.g., \cite{zygmund2002trigonometric}.
Thus, there exists an $m$ such that $c_0> \sum_{k\neq 0}|c_{k m}|$, and for this $m$, $\mathcal{T}_m(\theta) >0$ for all $\theta$.

An alternative argument can be drawn as follows. Denote by $L$ the torque's Lipschitz constant, i.e., $L=\inf\{\kappa \mid \left|\mathcal{T}(\theta+\Delta)-\mathcal{T}(\theta)\right|<\kappa|\Delta|\}$, for all $\theta,\Delta\in [0,2\pi]$. Then, $\mathcal{T}_m ( \theta)$ is also Lipschitz with Lipshitz constant $\leq L$. It is also periodic with period $2\pi/m$ and average $c_0$, which we assume positive. 
Let $\vartheta_0$ be such that $\mathcal T_m(\vartheta_0)=c_0$, which always exists since $\mathcal T_m$ is continuous. Then, over a period $\theta\in\left[\vartheta_0-\frac{\pi}{m},\vartheta_0+\frac{\pi}{m}\right]$,  $\mathcal{T}_m(\theta)$ takes
values in the interval $\left[c_0 - \tfrac{L\pi}{m},c_0 + \tfrac{L\pi}{m}\right]$. 
Thus, if we take $m = \lceil \tfrac{L\pi}{c_0} \rceil$, that is, we take the smallest integer $m$  such that $m \geq  \tfrac{L\pi}{c_0}$, it follows that $\mathcal{T}_m(\theta)>0$ over the period, and hence for all $\theta$.

We note that the number $m = \lceil \tfrac{L\pi}{c_0} \rceil$ of the needed engines is tight when $\mathcal T(\theta)$ has the shape of a triangular wave with  slope $L$ and period $2\pi$.

\subsection*{Alternative analysis for the Stirling case}\label{bound_stirling}
We derive a condition for  three coupled Stirling engines ($m=3$) to suffice for sustained limit cycle operation.

Let $\epsilon = \frac{sr}{V_0}$ and consider the expansion of the dimensionless torque in terms of $\epsilon$,
\footnotesize
\begin{align*}
    \frac{\mathcal{T}^S (\theta )}{\zeta nRT_0}& =\epsilon\left( \frac{ (1 + \alpha\tfrac{\Delta T}{2T_0} \sin(\theta) )}{1 + \epsilon(1-\cos\theta)}-\frac{p_0 V_0}{\zeta n R T_0}\right) \sin\theta  \\
    & =\epsilon \left(1+\alpha \frac{\Delta T}{2T_0}\, \sin\theta - \frac{p_0V_0}{\zeta n RT_0}\right)\sin\theta  + \mathcal{O}(\epsilon^2) \\
    &= \epsilon\, \frac{\alpha \Delta T}{4 T_0} + \epsilon \, \left(1-\frac{p_0V_0}{\zeta n R T_0}\right)\sin\theta + \epsilon \frac{\alpha \Delta T}{4 T_0}\cos(2\theta) + \mathcal{O}(\epsilon^2).
\end{align*} 
\normalsize Note that the two first harmonics vanish when coupling three Stirling engines, leaving only the constant term and higher order terms in $\epsilon$. Therefore, as long as 
$$
\frac{\alpha \Delta T}{4T_0} \gg \epsilon,
$$ 
three engines are enough to ensure that the torque is sign-definite. The resulting system will gyrate at approximately constant angular velocity
\begin{equation*}
 \langle \omega\rangle\approx \frac{\alpha \Delta T  \zeta n R}{4\Gamma}\epsilon.
\end{equation*}

\subsection*{Alternative analysis for the Brownian case}\label{bound_brownian}
In analogy with the Stirling engine, we expand the dimensionless torque for the Brownian gyrating engine in the dimensionless parameter $\beta$. This parameter controls  the variation of the capacitance and, by expanding around zero, we assume that this variation is small. That is, we assume that our system is within the linear response regime. The expansion gives
\begin{align*}
    \frac{\mathcal{T}^B(\theta)}{k_B T_0} &= 
    f_1(\theta) \beta +  f_2(\theta) \beta^2  +\mathcal{O}(\beta^3),
\end{align*}
where $f_1(\theta)$ depends on $\theta$ through terms linear in $\sin(\theta)$ and $\cos(\theta)$, while $f_2(\theta)$ contains second harmonics and a constant term.
Terms independent of $\beta$  vanish, since for $\beta=0$ energy cannot be extracted from the system. Therefore, up to second order in $\beta$, the only term that contributes to the average torque is $f_2(\theta)\beta^2$, whose average value over a cycle can be computed to be
\begin{equation*}
   \frac{1}{k_B T_0}\int_0^{2\pi}\mathcal{T}^B(\theta)d\theta\approx \frac{\beta^2}{2\pi} \int_0^{2\pi}f_2(\theta) d\theta = \frac{\sqrt{3}\Delta T}{64 T_0}\beta^2.
\end{equation*}

Consequently, if two engines are coupled, the first order term in $\beta$ vanishes, whereas, if three engines are coupled the remaining terms are of third order or higher.
Thus, provided
\begin{equation*}
    \frac{\sqrt{3} \Delta T}{64 T_0} \gg \beta,
\end{equation*}
the constant term dominates over higher order terms and a globally attractive limit cycle operation is present for the three coupled engines. In that case, the average angular velocity can be approximated by
\begin{equation*}
    \langle \omega \rangle  \approx \frac{\sqrt{3} k_B \Delta T}{64\Gamma}\beta^2.
\end{equation*}

\section{Optimizing phase difference}\label{optimization}
We now expand on the point raised in Section \ref{sec:equalizing}, that phase differences between coupled engines may be optimized to minimize the variation of the effective torque. Doing so, for two coupled engines, amounts to solving the following optimization problem
\begin{equation*}
    \min_{\theta_0} \frac{1}{2\pi}\int_0^{2\pi} \left(\tfrac{\mathcal{T}(\theta) + \mathcal{T}(\theta + \theta_0)}{2}-\langle{\mathcal T}\rangle\right)^2d\theta. 
\end{equation*}
Due to the periodicity of $\mathcal T$, the problem reduces to minimizing the integral of the product $\mathcal{T}(\theta)\mathcal{T}(\theta+ \theta_0)$ over a cycle. 

We bring in the Fourier series expansion \eqref{eq:expansion}, written for the terms in  this product, and consider the partial derivative of the integral with respect to $\theta_0$ so as to obtain the first order condition for optimality
\begin{equation*}
    -\sum_{k=1}^\infty k (a_k^2+b_k^2)\sin(k\theta_0)=0.
\end{equation*}
We see that $\theta_0 = n\pi,n\in \mathbb{N}$ are solutions and thus, potential extrema. Minimality hinges on the second derivative, which suffices to be strictly positive, i.e.,
\begin{equation*}
    -\sum_{k=1}^\infty k^2 (a_k^2 + b_k^2 )\cos(k\theta_0) >0.
\end{equation*}
It is clear that $\theta_0 = 0$ corresponds always to a maximum, while $\theta_0 = \pi$ may correspond to a maximum, a minimum, or be inconclusive, depending on the torque profile as a function of $\theta$. For instance, assuming $b_1$ and $a_2$ are the only nonzero terms in the Fourier expansion, as is the case of the Stirling engine (up to second order approximation in $\epsilon$), $\theta_0=\pi$  corresponds to a maximum as long as $b_1^2<4a_2^2$. In this case, there are two other extrema at 
\begin{equation*}
    \theta_0^{\rm opt} = \pm\text{acos}\left(-\frac{b_1^2}{4a_2^2}\right),
\end{equation*}
which are in fact minima.
All in all, the optimal phase difference for two coupled Stirling engines is equal to $\theta_0^{\rm opt}=\pi$ when \[
\left(1-\frac{p_0V_0}{\zeta n R T_0}\right)^2 >\left(\frac{\alpha\Delta T}{2T_0}\right)^2,
\]
and it is
\begin{equation*}
    \theta_0^{\rm opt} =\pm\text{acos} \left( - 4 \left(\frac{\zeta n R T_0 - p_0V_0}{\zeta n R \alpha \Delta T}\right)^2\right),
\end{equation*}
otherwise.
For the parameters used in this paper, it follows that the optimal phase is exactly $\pi$.

\vspace{0.5cm}
\section{Parameters used}\label{parameters}
The parameters we have used in the different numerical experiments are specified in Table \ref{tab:stirling}. Note that, for proper comparison, $\Gamma$ has been chosen such that $\log_{10}(\mathcal{I}/\mathcal{I}_0)=2$ both for the Stirling and the Brownian gyrating engines in Fig.~\ref{fig:stirling_DTomega_log2} and \ref{fig:brownian_DTomega_log2}, respectively.

\begin{table}[h]
\begin{tabular}{@{}ccc@{}}
\hline
\multicolumn{3}{c}{Stirling engine problem}
\\ \hline
     Parameter & Value & Units \\ \hline
     $s$     &   $71 $   &    $\text{mm}^2$   \\
     $r$     &   $ 3.5$   &  $\text{mm}$     \\
     $\zeta$     &  $0.94$     &    - \\
     $p_0$     &  $101.3$     &    $\text{kPa}$   \\
    $n$      &    $0.00185$   &  $\text{mol}$     \\
     $R$     &    $8.314$   &   $\text{J}\, \text{K}^{-1}\, \text{mol}^{-1}$    \\
     $T_\text{top}$     &    $297.15$   & $\text{K}$      \\
      $\alpha$    &    $0.17$   &  -     \\
    $V_0$      &   $44 900$    &   $\text{mm}^3$    \\
    $\mathcal{I}$     &  $10^{-1}$ to $10^{-8}$ (fig.~\ref{fig:stirling_omegaIGamma123})     &$\text{kg}\, \text{m}^2$  \\
    $\mathcal{I}$&$5.7\cdot 10^{-5}$ (fig.~\ref{fig:stirling_DTomega_log2}) & $\text{kg}\, \text{m}^2$ \\
    $\Delta T$     &  $10$ (fig.~\ref{fig:stirling_omegaIGamma123})     & $\text{K}$ \\
    $\Delta T$&$0$ to $15$ (fig.~\ref{fig:stirling_DTomega_log2})&$\text{K}$ \\
    $\Gamma$      &   $4.38\cdot 10^{-6}$    & $\text{kg}\, \text{m}^2 \, \text{s}^{-1}$      \\  \hline
    \multicolumn{3}{c}{Brownian gyrator problem} \\ \hline
     Parameter & Value & Units \\ \hline
     $C_0$ & $2$ & $\text{mF}$\\    
     $\beta$ & $0.1$ & - \\     
     $R_1,R_2$ & $1$ & $\Omega$ \\    
     $T_1$ & $200$& $\text{K}$ \\    
     $k_B$ & $1.38\cdot 10^{-23}$ & $\text{kg}\,\text{m}^2 \,  \text{s}^{-2}\, \text{K}^{-1}$\\    
    $\mathcal{I}$ &$10^{-12}$ to $10^{-19}$ (fig.~\ref{fig:brownian_omegaI_log2}) &  $\text{kg}\, \text{m}^2$\\ 
    $\mathcal{I}$&$5\cdot 10^{-16}$ (fig.~\ref{fig:brownian_DTomega_log2})&  $\text{kg}\, \text{m}^2$\\
    $\Delta T$ & $10$ (fig.~\ref{fig:brownian_omegaI_log2})&  $\text{K}$\\ 
    $\Delta T$ &$0$ to $15$ (fig.~\ref{fig:brownian_DTomega_log2})&$\text{K}$\\
     $\Gamma$ & $4.32\cdot 10^{-22}$ &$\text{kg}\, \text{m}^2\, \text{s}^{-1}$ \\ 
    \hline
\end{tabular}
\centering
\caption{Parameters used}
\label{tab:stirling}
\end{table}

\bibliography{apssamp}

\end{document}